\documentclass[tradiabstract]{aa} % for the abstract without structuration 

\usepackage{graphicx}
\usepackage{txfonts}

\usepackage{epsf}
\usepackage{rotating}
\usepackage{graphics}

\def \sophie{{\it SOPHIE}}
\def \MJ{M$_{\mathrm{Jup}}$}

\def \kms{km\,s$^{-1}$}
\def \ms{m\,s$^{-1}$}
\def \1s{$1\,\sigma$}
\def \kid{$\chi^2$}
\def \t0{T$_0$}

\def \cible{HD\,9446}
\def \cibleb{{\cible}b}
\def \ciblec{{\cible}c}

\begin{document}

   \title{The SOPHIE search for northern extrasolar planets\thanks{Based 
            on observations collected with the {\it SOPHIE}  spectrograph on the 1.93-m telescope at
            Observatoire de Haute-Provence (CNRS), France, by the {\it SOPHIE}  Consortium 
            (program 07A.PNP.CONS).}}
            
   \subtitle{II. A multi-planet system around \cible}

   \author{G.~H\'ebrard\inst{1}, 
                 X.~Bonfils\inst{2},
                 D.~S\'egransan\inst{3},
                 C.~Moutou\inst{4}, 
                 X.~Delfosse\inst{2},
	        F.~Bouchy\inst{1,5}, 
                 I.~Boisse\inst{1},
                 L.~Arnold\inst{5},
                 M.~Desort\inst{2},
                 R.~F.~D\'{\i}az\inst{1},
                 A.~Eggenberger\inst{2},  
                 D.~Ehrenreich\inst{2},
                 T.~Forveille\inst{2}, 
                 A.-M.~Lagrange\inst{2},
                 C.~Lovis\inst{3},
                 F.~Pepe\inst{3},
                 C.~Perrier\inst{2}, 
                 F.~Pont\inst{6}, 
                 D.~Queloz\inst{3},
                 N.~C.~Santos\inst{3,7},
	        S.~Udry\inst{3}, 
                 A.~Vidal-Madjar\inst{1}
}

   \institute{
%1
Institut d'Astrophysique de Paris, UMR7095 CNRS, Universit\'e Pierre \& Marie Curie, 
98bis boulevard Arago, 75014 Paris, France 
\and
%2
Universit\'e J.~Fourier (Grenoble 1)/CNRS, Laboratoire d'Astrophysique de Grenoble (LAOG, UMR5571), France
\and
%3
Observatoire de Gen\`eve,  Universit\'e de Gen\`eve, 51 Chemin des Maillettes, 1290 Sauverny, Switzerland
\and
%4
Laboratoire d'Astrophysique de Marseille, Universit\'e de Provence, CNRS (UMR 6110), 
BP 8, 13376 Marseille Cedex 12, France
\and
%5
Observatoire de Haute-Provence, CNRS/OAMP, 04870 Saint-Michel-l'Observatoire, France
\and
%6
School of Physics, University of Exeter, Exeter, EX4 4QL, UK 
\and
%7
Centro de Astrof{\'\i}sica, Universidade do Porto, Rua das Estrelas, 4150-762 Porto, Portugal
}

   \date{Received TBC; accepted TBC}
      
\abstract{We report the discovery of a planetary system around \cible, performed from 
radial velocity measurements secured with the spectrograph \sophie\ at the 193-cm 
telescope of the Haute-Provence Observatory during more than two years. At least two 
planets orbit this G5V, active star: 
\cibleb\ has a minimum mass of 0.7\,\MJ\ and a slightly eccentric orbit with a period of 30~days, 
whereas \ciblec\ has a minimum mass of 1.8\,\MJ\ and a circular orbit with a period of 193~days.
As for most of the known multi-planet systems, the \cible-system presents a hierarchical 
disposition, with a massive outer planet and a lighter inner planet.}


\authorrunning{H\'ebrard et al.}
\titlerunning{A multi-planet system around \cible}

\maketitle

%________________________________________________________________

\section{Introduction}
\label{sect_intro}

Among the more than 400 exoplanets known so far, most of them have been 
discovered from the reflex motion they cause to their host-star, which can be 
detected from stellar radial velocity wobble.  Thus, accurate radial velocity 
measurements remain a particularly efficient and powerful technic 
for research and characterization of exoplanetary systems. They allow the 
statistic of systems to be extended by completing the minimum mass-period 
diagram of exoplanets, in particular towards lower masses and longer 
periods, as the measurement accuracy improves.

Together with the advent of the new \sophie\ spectrograph at the 1.93-m telescope 
of Haute-Provence Observatory (OHP), France, the \sophie\ Consortium (Bouchy et 
al.~\cite{bouchy09}) started in late-2006 a large observational program of 
exoplanets search and characterization, using the radial velocity technique. 
In the present paper we announce the discovery of two exoplanets around 
\object{\cible}, from radial velocity measurements secured as part of the 
second sub-program of the \sophie\ Consortium. This sub-program 
is a giant-planet survey on 
a volume-limited sample around 2000 FGK stars, requiring moderate accuracy, 
typically in the range $5-10$\,\ms\ (Bouchy et al.~\cite{bouchy09}). Its goal 
is to improve the statistics on the exoplanet parameters and their hosting stars by 
increasing the number of known Jupiter-mass planets, as well as offer a chance 
to find new transiting giant planets in front of bright stars. \sophie\  sub-program-2 
data were already used to report the detection of several planets (Da Silva et 
al.~\cite{dasilva08}, Santos et al.~\cite{santos08}, Bouchy et al.~\cite{bouchy09}) 
and to study stellar activity (Boisse et al.~\cite{boisse09a}). This sub-program also 
aims at following up transiting giant exoplanets; this allowed spectroscopic 
transits to be observed (Loeillet et al.~\cite{loeillet08}), including the detection of the 
two first cases of spin-orbit misalignment, namely XO-3b 
(H\'ebrard et al.~\cite{hebrard08}) then HD\,80606b (Moutou et al.~\cite{moutou09}, 
Pont et al.~\cite{pont09}), simultaneously with the discovery of the transiting nature 
of the planet in this last case.

The \sophie\ observations of \cible\ that allow the detection of two new planets 
are presented in section~\ref{sect_observations}. We derive and 
discuss the stellar and planetary properties in sections~\ref{sect_stellar_properties} 
and~\ref{sect_planetary_system} respectively, and conclude in 
section~\ref{sect_conclusion}.

\section{Observations}
\label{sect_observations}

We observed \cible\ with the OHP 1.93-m telescope and \sophie, which is a 
cross-dispersed, environmentally stabilized echelle spectrograph dedicated to 
high-precision radial velocity measurements (Perruchot et al.~\cite{perruchot08}, 
Bouchy et al.~\cite{bouchy09}). Observations were secured in \textit{high-resolution} 
mode, 
allowing the resolution power 
$\lambda/\Delta\lambda=75000$ to be reached. 
The spectra were obtained in three seasons, from November~2006 to March~2009.
Depending of variable atmospheric conditions, the exposure times ranged between 
3 and 18~minutes and signal-to-noise ratios per pixel at 550~nm between 32 and 94, 
with typical values of 5.5~minutes and 55, respectively. Exposure time and signal-to-noise 
ratio were slightly greater during the first season of observation. Three exposures performed 
through too cloudy conditions were excluded from the final dataset, that 
includes 79 spectra. The total exposure time is about 7~hours.

The spectrograph is fed by two 
optical fibers, the first one being used for starlight. During the first season the second 
\sophie\ entrance fiber was fed by a thorium lamp for simultaneous wavelength calibration. 
Thereafter we estimated that wavelength calibration performed with a $\sim2$-hour 
frequency each night (allowing interpolation for the time of the exposure) was sufficient, 
and that the instrument was stable enough to avoid simultaneous 
calibration for this moderate-accuracy program. So for the second and third seasons, 
no simultaneous thorium calibration were performed, avoiding pollution of the 
first-entrance spectrum by the calibration light. The second entrance fiber
was instead put on the sky; this allowed us to check that none 
of the spectra were significantly affected by sky background pollution, especially due 
to~moonlight. 

We used the \sophie\ pipeline (Bouchy et al.~\cite{bouchy09}) to 
extract the spectra from the detector images, 
cross-correlate them with a G2-type numerical mask, 
then fit the cross-correlation functions (CCFs) by Gaussians to get the radial velocities 
(Baranne et al.~\cite{baranne96}, Pepe et al.~\cite{pepe02}).
Each spectrum produces 
a clear CCF, with a  $8.47 \pm 0.05$~km\,s$^{-1}$ full width at half 
maximum and a contrast representing $38.0 \pm 0.8$~\%\ of the continuum.
Only the 33 first spectral orders of the 39 available ones 
were used for the cross correlation; this 
allows the full dataset to be reduced with~the same procedure, as the last red orders 
of the \cible~spectra obtained during the first season were polluted by 
argon spectral lines of the simultaneous wavelength-calibration.

The derived radial velocities are reported in Table~\ref{table_rv}. The accuracies 
are between 5.4 and 9.7\,\ms, typically around 6.5\,\ms. This includes photon noise 
(typically $\sim3.5$~\ms), wavelength calibration ($\sim3$~\ms), and 
guiding errors ($\sim4$~\ms) that produce motions of the input image within the 
fiber (Boisse et al.~\cite{boisse09b}). These computed uncertainties do not 
include any ``jitter'' due to stellar activity (see below).

\begin{table}[h]
  \centering 
  \caption{Radial velocities of \cible\ measured with {\it SOPHIE} (full table available electronically).}
  \label{table_rv}
\begin{tabular}{ccc}
\hline
\hline
BJD & RV & $\pm$$1\,\sigma$  \\
-2\,400\,000 & (km\,s$^{-1}$)  & (km\,s$^{-1}$)  \\
\hline
54043.4849  &  21.7152  &  0.0063  \\
54045.4850  &  21.7451  &  0.0055  \\
54047.4598  &  21.7747  &  0.0056  \\
54048.5033  &  21.7969  &  0.0054  \\
\ldots & \ldots & \ldots \\
\ldots & \ldots & \ldots \\
54889.2599  &  21.7201  &  0.0073  \\
54890.2628  &  21.7409  &  0.0064  \\
54893.2986  &  21.7222  &  0.0090  \\
54894.2961  &  21.7074  &  0.0097  \\
\hline
\end{tabular}
\end{table}

\section{Stellar properties of \cible}
\label{sect_stellar_properties}

We used the 50 \sophie\ spectra secured without simultaneous thorium exposure 
to obtain an averaged spectrum, and we managed a spectral analysis from it. 
Table~\ref{table_stellar_parameters} summarizes the stellar parameters. 
According to the SIMBAD database, \cible\ (HIP\,7245, BD+28\,253) is a $V=8.35$, 
high proper-motion G5V star. Its Hipparcos parallax ($\pi=19.92\pm1.06$~mas) 
implies a distance of $53 \pm 3$~pc. The Hipparcos color is $B-V=0.680 \pm 0.015$
(Perryman et al.~\cite{perryman97}).

\begin{table}[h]
  \centering 
\caption{Adopted stellar parameters for \cible.}
  \label{table_stellar_parameters}
\begin{tabular}{lc}
\hline
\hline
Parameters  & Values \\ 
\hline
$m_v$                		&	$8.35$ 			\\ 
Spectral~type        		&	G5V				\\ 
$B-V$          			&	$0.680 \pm 0.015$ 	\\ 
Parallax [mas]			&	$19.92\pm1.06$	\\ 
Distance [pc]     		&	$53 \pm 3$ 		\\ 
$v\sin i_\star$ [\kms]		&	$4 \pm 1$			\\ 
$\log{R'_\mathrm{HK}}$	&	$-4.5 \pm 0.1$		\\ 
${\rm [Fe/H]}$ 			&	$0.09 \pm 0.05$	\\ 
$T_{\rm eff}$ [K]		&	$5793 \pm 22$		\\ 
$\log g$ [cgi] 			&	$4.53 \pm 0.16$	\\ 
Mass~$[\rm{M}_{\odot}]$		&	$1.0 \pm 0.1$	\\ 
Radius~$[\rm{R}_{\odot}]$		&	$1.0$	\\ 
Luminosity~$[\rm{L}_{\odot}]$		&	$1.1$	\\ 
\hline
\end{tabular}
\end{table}

From spectral analysis of the \sophie\ data using the method presented in Santos et 
al.~(\cite{santos04}), we derived the temperature $T_{\rm eff} = 5793 \pm 22$~K, the 
gravity $\log g = 4.53 \pm 0.16$, 
${\rm [Fe/H]} = +0.09 \pm 0.05$, and $M_* = 1.0 \pm 0.1 \,\rm{M}_{\odot}$.
The 10\,\%\ uncertainty on the stellar mass is an estimation, systematic
effects being difficult to quantify (Fernandes \& Santos~\cite{fernandes04}). 
We derive a projected rotational velocity 
$v\sin i_\star = 4 \pm 1$~\kms\ from the parameters of the CCF using 
the calibration of Boisse et al.~(in preparation), 
%Boisse et al.~(\cite{boisse10}),
which is  similar to that 
presented by Santos et al.~(\cite{santos02}). We also obtained 
${\rm [Fe/H]} = +0.12 \pm 0.10$ from the CCF, which agrees with,~but is less accurate 
than the metallicity obtained from our spectral~analysis. 

\begin{figure}[h] 
\begin{center}
\vspace{1cm}
\includegraphics[scale=0.52]{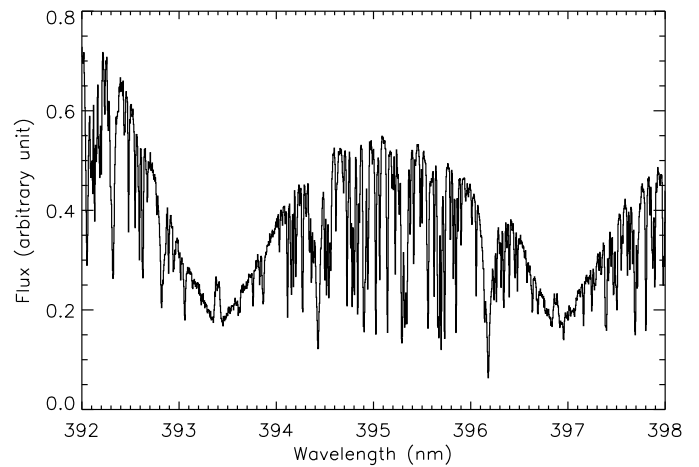}
\caption{H and K \ion{Ca}{ii} lines of \cible\ on the averaged \sophie~spectra. 
Chromospheric emissions are detected, yielding a $\log{R'_\mathrm{HK}}=-4.5 \pm 0.1$.}
\label{fig_caII}
\end{center}
\end{figure}

The cores of the large \ion{Ca}{ii} absorption lines of \cible\ show small emissions
(Fig.~\ref{fig_caII}), which are the signature of an active chromosphere. Such stellar activity 
would imply a significant ``jitter'' on the stellar radial velocity measurement. The 
level of the \ion{Ca}{ii} emission corresponds to $\log{R'_\mathrm{HK}}=-4.5$ 
with a  $\pm 0.1$ dispersion  
according to the \sophie\ calibration (Boisse et al.~in preparation). 
%(Boisse et al.~\cite{boisse10}).
For a G-type star with this 
level of activity, Santos et al.~(\cite{santos00}) predict a dispersion of the order of 10 to 
20~\ms\ for the stellar jitter. 
According to Noyes et al.~(\cite{noyes84}) and Mamajek \& Hillenbrand~(\cite{mamajek08}), 
this level of activity implies a stellar rotation period $P_\mathrm{rot}\simeq10$~days. 
This agrees with our $v\sin i_\star$ measurement, which translates into 
$P_\mathrm{Rot} < 17$~days (Bouchy et al.~\cite{bouchy05}), depending on the 
unknown inclination $i_\star$ of the stellar rotation~axes.

\section{A planetary system around \cible}
\label{sect_planetary_system}

\begin{figure}[h] 
\begin{center}
\includegraphics[scale=0.44]{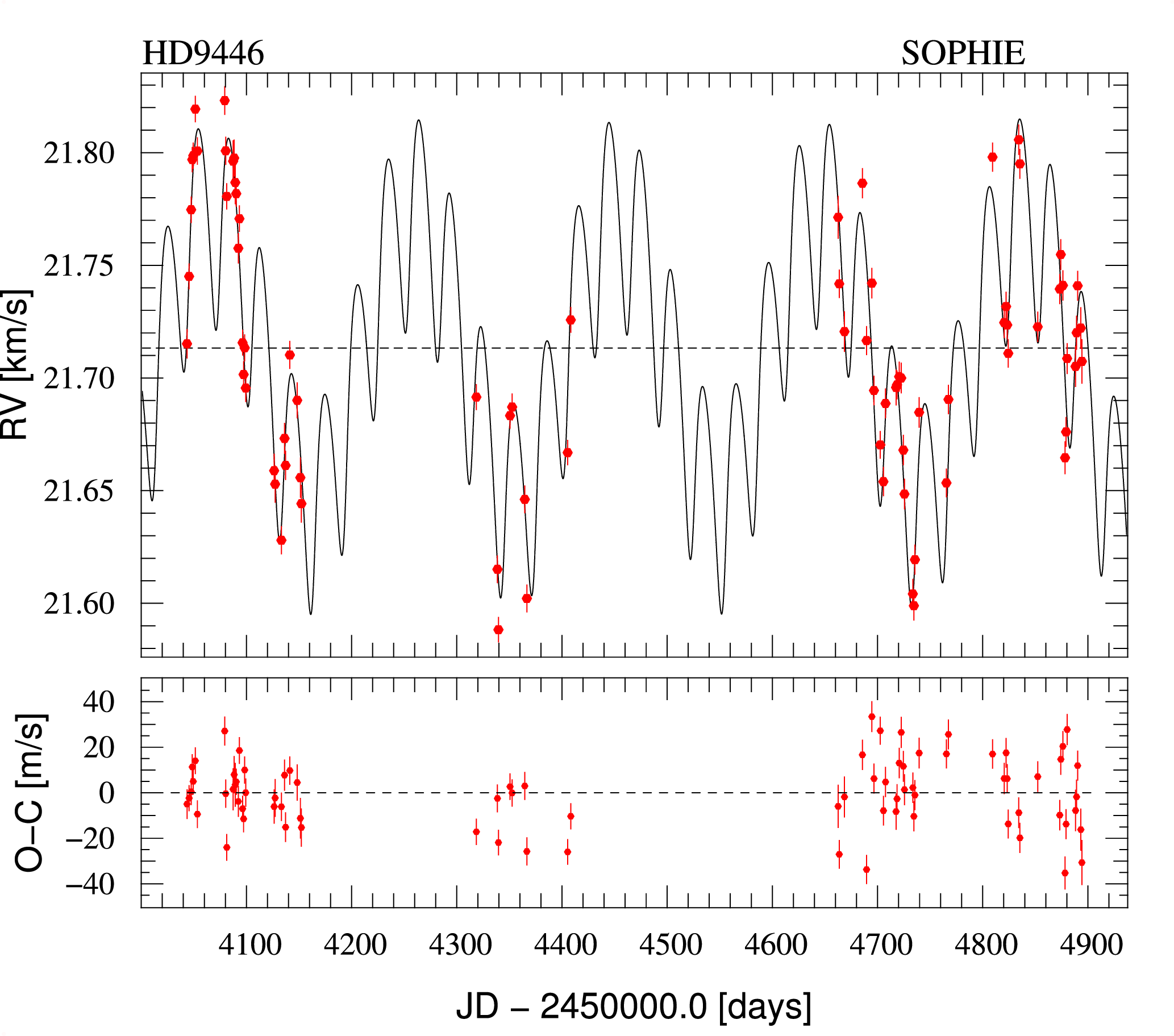}
\caption{\textit{Top:} Radial velocity  {\it SOPHIE} measurements of \cible\ 
as a function of time, and Keplerian fit with two planets. 
The orbital parameters corresponding to this 
fit are reported in Table~\ref{table_parameters}. 
\textit{Bottom:} Residuals of the fit with 1-$\sigma$~error bars.}
\label{fig_omc}
\end{center}
\end{figure}

\begin{figure}[h] 
\begin{center}
\vspace{1cm}
\includegraphics[scale=0.58]{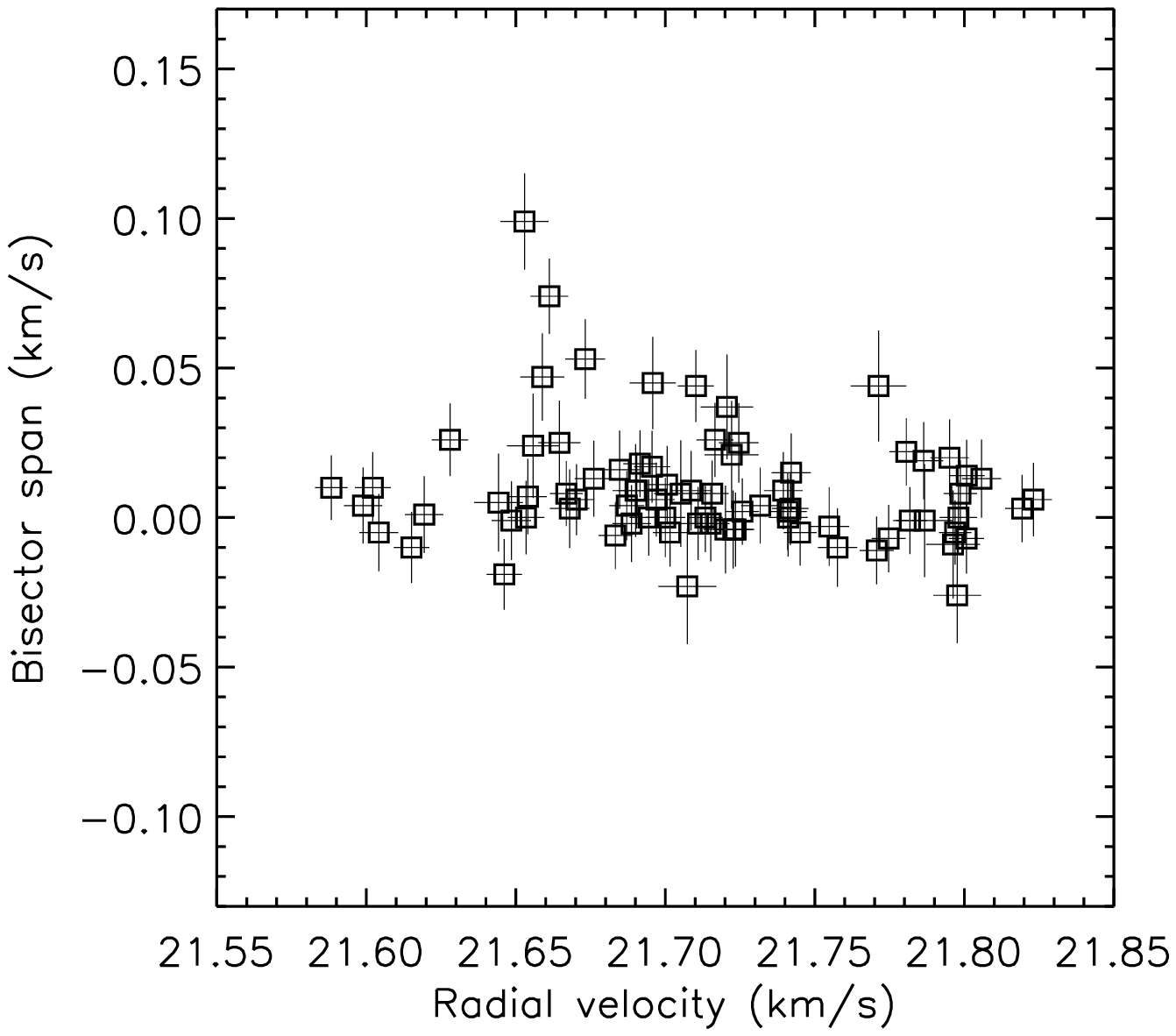}
\includegraphics[scale=0.58]{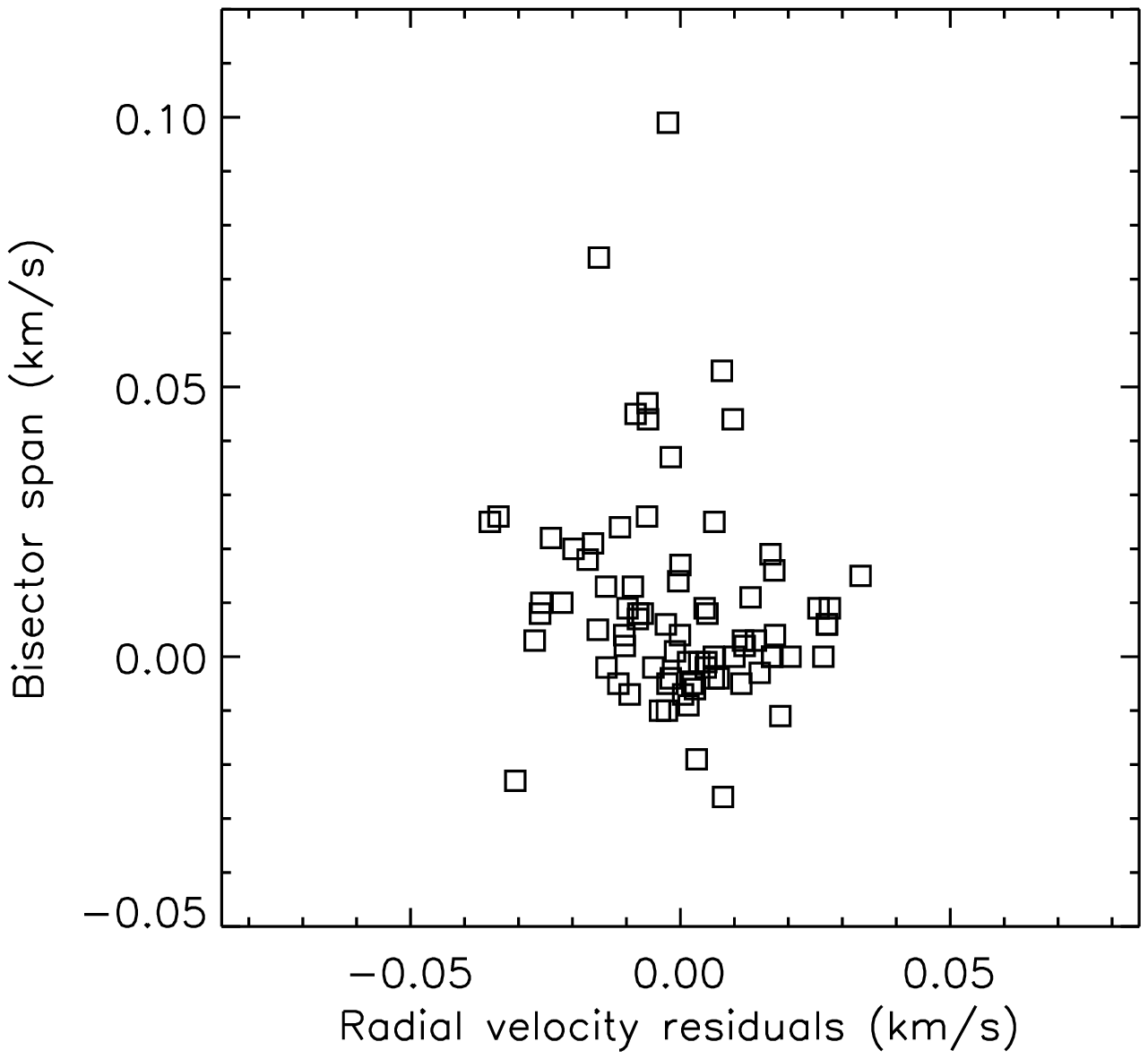}
\caption{Bisector span as a function of the radial velocity (top) and the radial 
velocity residuals after the 2-planet fit (bottom). 
For clarity, error bars are not plotted in the bottom panel.  
The ranges have the same extents in the $x$- and $y$-axes on both panels.
}
\label{fig_bis}
\end{center}
\end{figure}

The \sophie\ radial velocities of \cible\ are plotted in Fig.~\ref{fig_omc}. 
Spanning more than two years, they 
show clear variations of the order of 200~\ms, implying a dispersion 
$\sigma_{\mathrm{RV}}=58$\,\ms; this is well over the expected stellar jitter due 
to chromospheric activity (10 to 
20~\ms, see above).  In addition, the bisectors of the CCF are stable 
(Fig.~\ref{fig_bis}, upper panel), 
showing dispersion of the order of $\sigma_{\mathrm{BIS}} = 20$~\ms, 
well below that of the radial velocities. An anticorrelation between the bisector and the 
radial velocity is usually the signature of radial velocity variations induced by stellar activity
(see, e.g., Queloz et al.~\cite{queloz01}, Boisse et al.~\cite{boisse09a}). The bisectors
are flat by comparison with the radial velocities, which 
suggests that the radial velocity variations are mainly due to Doppler shifts of the stellar lines 
rather than stellar profile variations. This leads to conclude that reflex motion due to 
companion(s) are the likely cause of the stellar radial velocity~variations.

These facts were known in late 2007, after two seasons of \sophie\ observations 
of \cible. A search of Keplerian fits then produced a solution with two Jupiter-like 
planets, on orbits of 30 and 190 days of period, with low eccentricities.
This solution was 
thereafter confirmed by the third season of observation; together with the ``flat''
bisectors, this provides a strong support to the two-planet interpretation of the 
radial velocity variations.

\begin{figure}[h] 
\begin{center}
\vspace{1cm}
\includegraphics[scale=0.43]{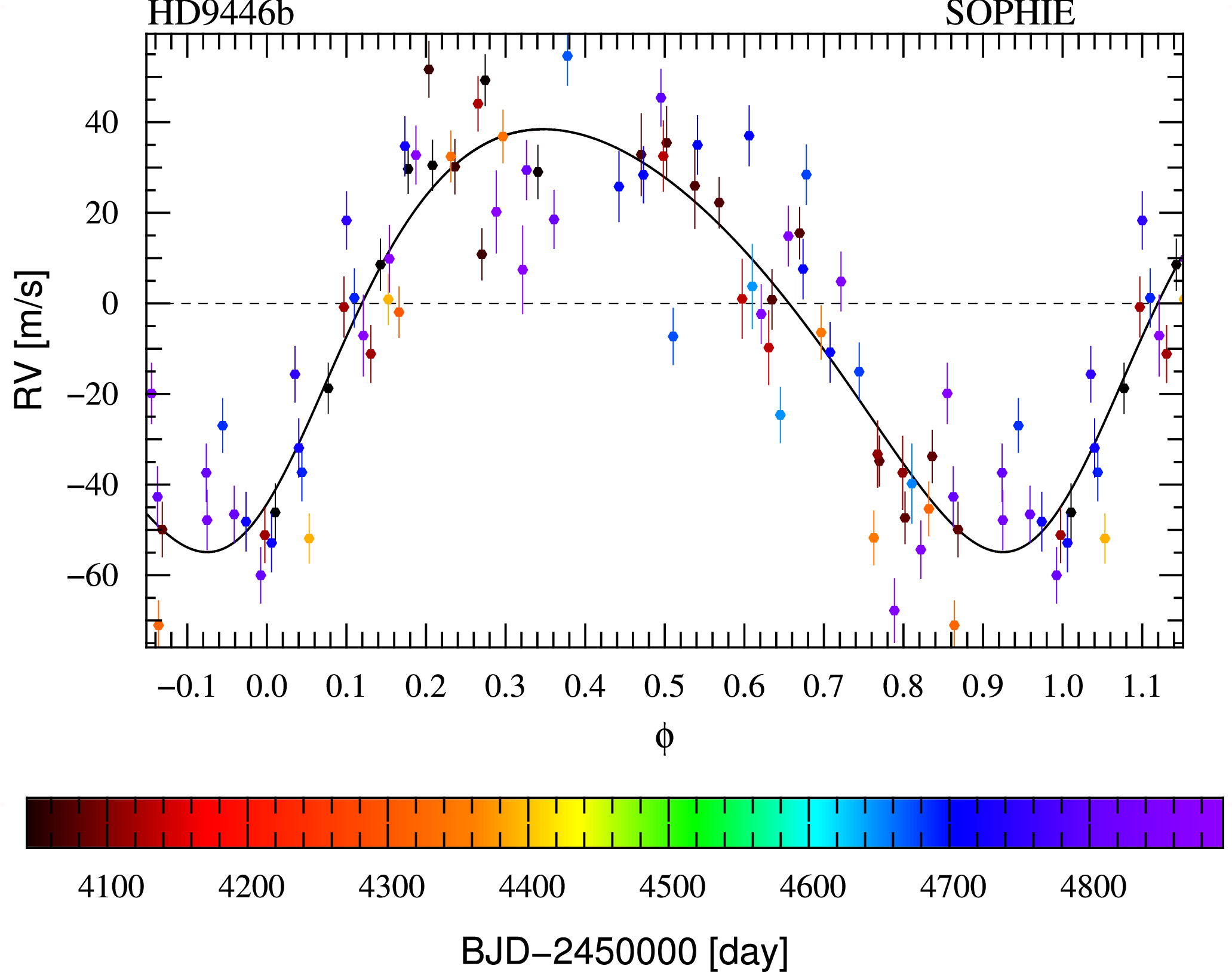}
\includegraphics[scale=0.43]{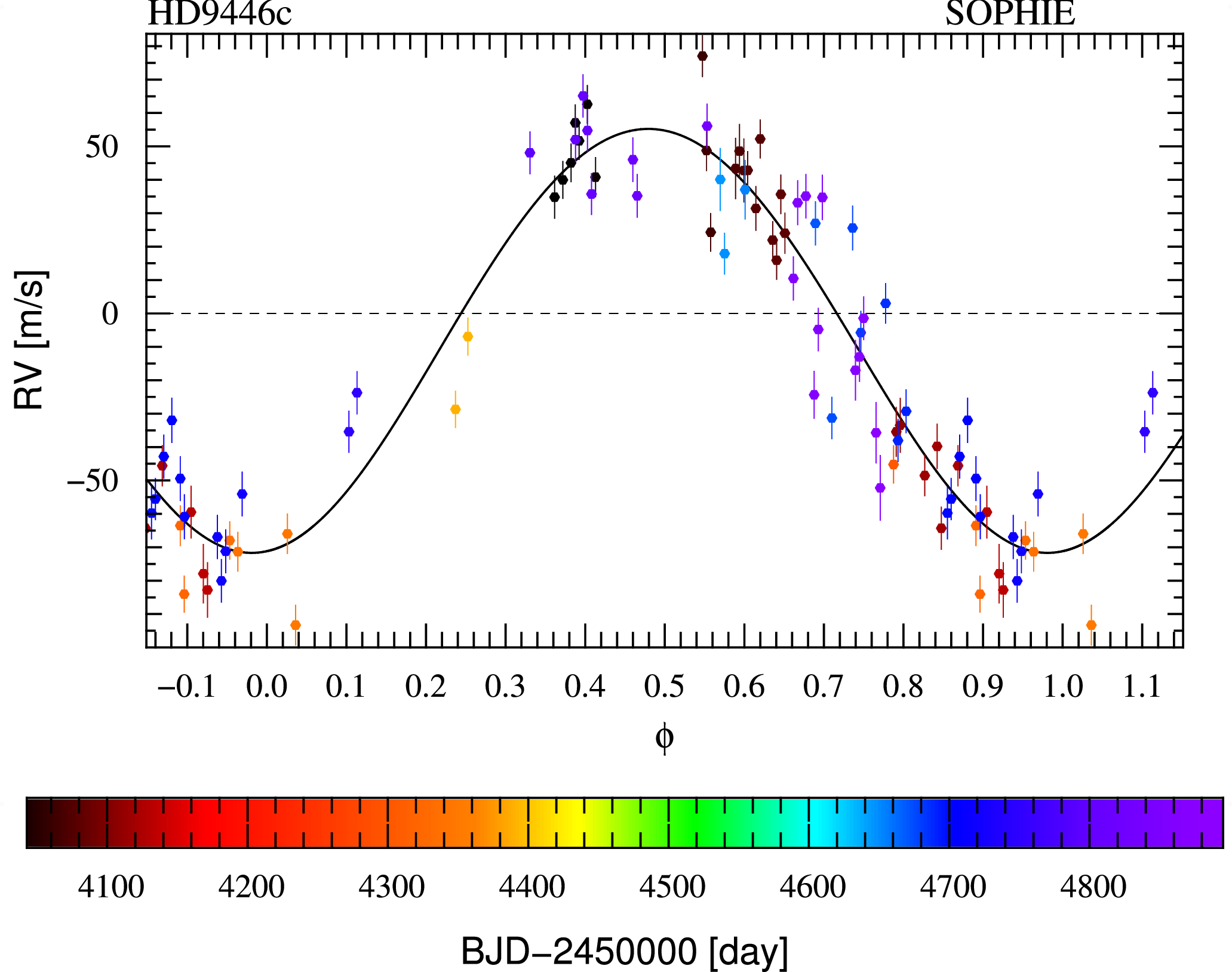}
\caption{Phase-folded radial velocity curves for \cibleb\ ($P=30$\,d, top) 
and \ciblec\ ($P=193$\,d, bottom) after removing the effect of the other planet.
The {\it SOPHIE} radial velocity measurements are presented with 
1-$\sigma$~error bars, and the Keplerian fits are the solid lines. 
Orbital parameters corresponding to the 
fits are reported in Table~\ref{table_parameters}.
The colors indicate the measurement dates.}
\label{fig_orb_phas}
\end{center}
\end{figure}

\begin{table}[h]
  \centering 
  \caption{Fitted orbits and planetary parameters for the \cible\ system, with 1-$\sigma$ error bars.}
  \label{table_parameters}
\begin{tabular}{lcc}
\hline
\hline
Parameters & \cibleb\ & \ciblec\ \\
\hline
$P$ 				[days]			& $30.052\pm 0.027$		&  $192.9\pm 0.9$	\\
$e$								& $0.20\pm 0.06 $			&  $0.06\pm 0.06 $	\\
$\omega$ 		[$^{\circ}$]			& $-145\pm30$				&  $-260\pm130$	\\
$K$				[\ms]				& $46.6\pm3.0$			&  $63.9\pm4.3$	\\
$T_0$ (periastron)	[BJD]			& $2\,454\,854.4\pm2.0$		&  $2\,454\,510\pm70$ 	\\
$M_\textrm{p} \sin i$	[M$_\mathrm{Jup}$]	& $0.70 \pm 0.06$$^\dagger$	&  $1.82 \pm 0.17$$^\dagger$	\\
$a$				[AU]				& $0.189 \pm 0.006$$^\dagger$	&  $0.654 \pm 0.022$$^\dagger$	\\
$V_r$ 			[\kms]			& \multicolumn{2}{c}{$21.715\pm0.005$} 	\\
$N$								& \multicolumn{2}{c}{79}	 		\\
reduced \kid						& \multicolumn{2}{c}{2.6}	 		\\ 
$\sigma_\mathrm{O-C}$		[\ms]		& \multicolumn{2}{c}{15.1}	 	\\	
Typical RV accuracy [\ms]   			& \multicolumn{2}{c}{6.5} \\
span 			[days]			& \multicolumn{2}{c}{851}                      \\
\hline
\multicolumn{3}{l}{$\dagger$: using $M_\star = 1.0\pm0.1$\,M$_\odot$}
\end{tabular}
\end{table}

Figs.~\ref{fig_omc} and~\ref{fig_orb_phas} show the final fit of the 851-day span 
\sophie\ radial velocities of \cible. This Keplerian model includes two planets 
without mutual interactions, which are negligible in this case 
(see Sect.~\ref{sect_conclusion}). 
All the parameters are free to vary during the fit.
The derived orbital parameters are 
reported in Table~\ref{table_parameters}, together with error bars, which were 
computed from  \kid\ variations and Monte~Carlo experiments. 

The inner planet, \cibleb, produces radial velocity variations with a semi-amplitude 
$K=46.6\pm3.0$~\ms, corresponding to a planet with a minimum mass 
$M_\textrm{p} \sin i   = 0.70 \pm 0.06$~M$_\mathrm{Jup}$ (assuming 
$M_\star = 1.0\pm0.1$\,M$_\odot$ for the host star). Its orbit has a period of 
$30.052\pm 0.027$~days, and is significantly non-circular ($e=0.20\pm 0.06$).
This period is longer than the stellar rotation period, as determined 
above from the $\log{R'_\mathrm{HK}}$ and the $v\sin i_\star$. 
A $P_\mathrm{rot}$-value of 30~days would corresponds to 
$v\sin i_\star < 2$~\kms, which is incompatible with our data.
The outer planet, \ciblec, yields a semi-amplitude $K=63.9\pm4.3$~\ms, corresponding 
to a planet with a projected mass $M_\textrm{p} \sin i = 1.82 \pm 0.17$~M$_\mathrm{Jup}$.
The orbital period is $192.9\pm 0.9$~days. This is about half a Earth-year, which made 
difficult a good phase coverage for the observations. As seen in the lower panel of 
Fig.~\ref{fig_orb_phas}, the rise of the radial velocity due to  \ciblec\ lacks of 
measurements, for orbital phases between 0.0 and 0.3. 
This implies significant uncertainties on the shape of the orbit.
Circularity can not be excluded ($e=0.06\pm 0.06$); furthermore, if the orbit actually 
is eccentric, there are nearly no constraints with the present dataset on  
the orientation of the ellipse with respect to the line of sight. The resulting error bars
on the longitude $\omega $ of the periastron and on the time $T_0$ at periastron 
are thus large; there are however correlated, and the timing of a possible transit for this 
planet is better constrained than $T_0$ in Table~\ref{table_parameters}.
Our estimations of $v\sin i_\star$ and $P_\mathrm{rot}$ allow the constraint 
 $i_\star > 30^{\circ}$ to be put. So, if we assume a spin-orbit alignment for the 
 \cible-system,  $i_\star = i$ and $\sin i > 0.5$; this implies projected masses that 
 translate into actual masses clearly in the planetary range.

The reduced \kid\ of the Keplerian fit is 2.6, and the standard deviation of the residuals is 
$\sigma_\mathrm{O-C}=15.1$~m\,s$^{-1}$. This is better than the $58$-\ms\ dispersion of 
the original radial velocities, but this remains higher than the 6.5-\ms\ typical error bars on 
the individual measurements, suggesting an additional noise of $\sim 13.5$~\ms. Such 
dispersion is precisely in the range of the 10 to 20~\ms\ expected jitter for a 
G-type star with this level of activity (Sect.~\ref{sect_stellar_properties}). 
Stellar activity is thus likely to be the main cause of the remaining dispersion, 
as well as the $\sim20$-\ms\ dispersion of the bisectors. The residuals of the fits
do not show significant anticorrelation with the bisectors (Fig.~\ref{fig_bis}, lower panel), 
as it could be expected 
in such cases (see, e.g., Melo et al.~\cite{melo07}, Boisse et al.~\cite{boisse09a}); 
this is however at the limit of detection according to the error bars.
A few bisectors values are larger than the other ones. They could be due to a 
particularly active phase of the star, as they are localized in a 
short time interval (between late January and early February 2007). Excluding these outliers 
from the analysis does not significantly change the results.
Finally, as it can be seen on lower panel of Fig.~\ref{fig_omc}, the residuals are significantly 
less scattered during the first season than during the third one. 
This can be mainly explained by the 
higher signal-to-noise ratio reached with longer exposure times during the first season, 
as well as the simultaneous thorium calibration secured for the first~measurements. 

\begin{figure}[h] 
\begin{center}
\vspace{1cm}
\includegraphics[scale=0.59]{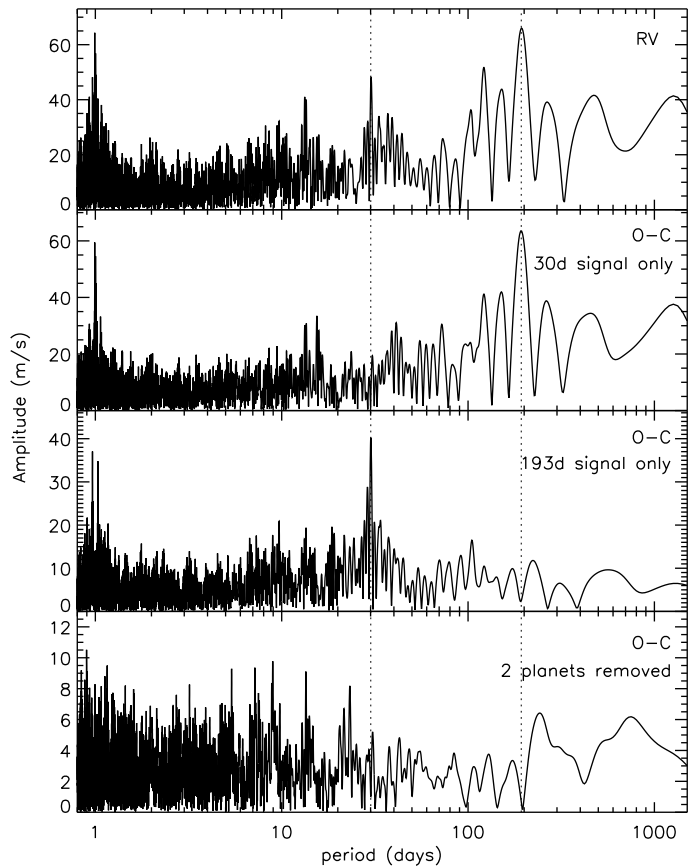}
\caption{Lomb-Scargle periodograms of the {\it SOPHIE} radial velocities. The upper 
panel shows the periodogram computed on the initial radial velocities, without any fit 
removed. The second and third panels show the periodograms computed on the residuals 
of the fits including \cibleb\ or \ciblec\ only, respectively. The bottom panel shows the 
periodogram after the subtraction of the 2-planet fit. 
The two vertical dotted lines show the periods of the two planets.
}
\label{fig_periodogram}
\end{center}
\end{figure}

Figure~\ref{fig_periodogram} shows Lomb-Scargle periodograms 
of the radial velocity measurements of \cible\ in four different cases: 
without any planet removed, with one 
or the other planet removed, and with both planets removed. A similar study was 
performed in the case of BD\,$-08^{\circ}2823$, another star with two detected planets 
(H\'ebrard et al.~\cite{hebrard09}).
In the upper panel of Fig.~\ref{fig_periodogram} that presents the 
periodogram of the raw radial velocity measurements of \cible,
periodic signals at $\sim30$~days and $\sim195$~days are clearly detected with 
peaks at those periods, corresponding to the two planets reported above, with 
the same amplitudes. The peak at $\sim1$~day corresponds to the aliases 
of all the detected signals, as the sampling is biased towards ``one point per night''.
A fourth, weaker peak is detected at $\sim13.3$~days. A Keplerian fit of this 
signal would provide a semi-amplitude $K\simeq11$~\ms, corresponding to a projected 
mass of 40 Earth masses. We do not conclude, however, that we detect a third, 
low-mass planet within the current data. 

Indeed, firstly this 13.3-day period is 
near the stellar rotation period ($\sim10$~days, Sect.~\ref{sect_stellar_properties}), 
so it could be at least partially due to stellar rotation. 
However, no significant peaks are detected at this period (nor at 30 days or 193 days) 
on the bisectors periodograms. 
We interpret this 13.3-d signal as being more likely due to aliases. 
In order to validate this, we constructed a fake radial velocity dataset with the same 
time sampling as our actual data, and that includes only the Keplerian model of the 
two planets found above. The periodogram of this fake dataset is almost identical 
as the one plotted in the upper panel Fig.~\ref{fig_periodogram}: it includes of course 
the two peaks corresponding to the periods of the two planets, but also the 
peak at 13.3~days. 

In addition to the one-day peak, the window function of our data 
shows a peak at $\sim14.3$~days, indicating that this interval 
is favored in our time sampling. The 13.3-d signal could thus be mainly due 
to the 14.3-day alias of the 192.9-day signal ($1/13.3 \simeq 1/14.3 + 1/192.9$).
On the second panel of Fig.~\ref{fig_periodogram} is plotted the periodogram 
of the residuals after subtraction of 
a fit including the 30-day-period planet only. The peak at 
193~days is visible, as well as these aliases at 1, 13.3, and 15.4~days 
($1/15.4 \simeq 1/14.3 - 1/192.9$). In the same manner, the third panel of 
Fig.~\ref{fig_periodogram} shows the periodogram of the residuals after a fit 
including the 193-day-period planet only. The peak at 193~days is no longer 
visible, and neither are the 
three aliases seen on the upper panel. This time the
peak at 30~days is visible, together with these two aliases due to the 1-day favored 
sampling, at 0.97 and 1.03~day.  The 
bottom panel of Fig.~\ref{fig_periodogram} shows the periodogram of the residuals 
after subtraction of the Keplerian fit including \cibleb\ and \ciblec. There are no remaining 
strong peaks on this periodogram; even the 1-day alias disappeared, showing that most 
of the periodic signals have been removed from the data. The remaining peaks are 
below 10~m/s amplitude, showing that the main part of the detected periodical signals 
in our data are due to the two planets. The remaining signal in the residuals are 
at the limit of detection according to our accuracy. As in addition the stellar rotation 
period is close to an alias of the signal of \ciblec, this makes tough any characterization 
of radial-velocity signal due to stellar activity, which is expected mainly at the 
stellar rotation~period. As the stellar jitter on the radial velocities is of 
the order of 10~\ms, this effect on the derived parameters of the two detected planets 
is~negligible.

The residuals of the measurements secured during the second observational 
season are preferentially negative (Fig.~\ref{fig_omc}, lower panel). This may 
suggest a possible additional component, with an orbital period of the same order 
or larger than the time span of our dataset (2.3 years). Such an additional planet 
could not be established with the available data. For a $\sim2$-yr period, 
the projected mass of such an hypothetic planet 
should be lower than one Jupiter mass. 
On the other hand on short periods, 
a hot-Jupiter is excluded in this system, the accuracy of our dataset being 
good enough to detect it if there were any. 
The data allow planets with masses larger than
0.3~\MJ\ and orbital periods shorter than 10~days to be excluded in the 
\cible\ system.

\section{Discussion}
\label{sect_conclusion}

The data we presented allow us to conclude there is a planetary system around 
\cible, with at least two Jupiter-like planets, on 30 and 193-day orbits. \cibleb\ has a 
projected mass slightly lower than Jupiter; it is on a 0.2-eccentricity orbit, showing that 
tidal effects were not strong enough to circularize it. \ciblec\ is at least 1.8 times more 
massive than Jupiter, and is on a nearly-circular orbit. The host-star of this system 
is slightly more metallic than the Sun, in agreement with the tendency found for stars 
harboring Jupiter-mass planets (see, e.g., Santos et al.~\cite{santos05}).

The mutual gravitational interactions between \cibleb\ and \ciblec\ are
weak. The inner planet is stabilized on its orbit by the strong
gravity of the star. Following Correia et al.~(\cite{correia05}), a simulation 
of the two orbits from the current solution was run for $10^6$~years, in order
to estimate their evolution from mutual interactions. 
This shows no significant changes in the eccentricities, which 
remain in the ranges [$0.18-0.23$] and [$0.03-0.075$], for \cibleb\ and \ciblec\ respectively.
Therefore this system is stable for $10^6$~years, and it seems to be stable for longer time 
scales also.
We estimated the order of magnitude of the potential transit timing variations
due to those weak mutual interactions, if any of the planets of the system does transit.
For that purpose we performed another 3-body simulation of the 
system, assuming the masses of the planets are equal to the minimum masses 
and that the orbits are coplanar. We employed the Burlisch-Stoer algorithm 
implemented in the Mercury6 package (Chambers~\cite{chambers99}) 
and integrated the 
system for 2000 days -i.e. around 10 orbits of the exterior planet. We found 
that the interaction between the planets produces variations in the central 
time of transits with small amplitude, that does not exceed 0.4~second for any 
of the two~bodies.

No photometric search for transits have been managed for \cible; depending on the 
unknown inclination  $i$ of the orbit, the transit probability for \cibleb\ and \ciblec\ are 
about 2\,\% and 1\,\%, respectively. There are more than 200 exoplanets detected 
from radial velocity surveys with orbital periods longer than 50 days, so with transit 
probabilities of the level of the percent. Only one is known to transit, namely HD\,80606b
(Moutou et al.~\cite{moutou09}). It is likely that at least one or two more of these known 
long-period exoplanets are actually transiting as seen from the Earth. Their search 
is challenging, as the times of the possible transits are not known accurately, especially 
a few years after the securement of the radial velocity~data. 

Among the more than 400 exoplanets discovered so far, almost 25\,\%\ are located in the 
$\sim40$ known multiple-planet systems. Most of them have been detected from radial 
velocity measurements. 
Additional planetary companions around \cible\ can not be detected with the available 
data besides the two planets reported, but they are of course 
possible, as multiple-planet systems are common. 
For example HD\,155358 
(Cochran et al.~\cite{cochran07}) has a Jupiter-mass planet on an orbit similar to \ciblec, 
and another planet with a 530-day orbital period, or HD\,69830 (Lovis et al.~\cite{lovis06})
have two Neptune-mass planets on orbits similar to those of the two detected planets 
of \cible, and a third one on a 8.7-day orbit. More data are needed, and 
the monitoring of \cible\ should thus be maintained.
As low-mass planets are preferentially found 
in multiple planetary systems (see, e.g., McArthur et al.~\cite{mcarthur04}, 
Pepe et al.~\cite{pepe07}, Mayor et al.~\cite{mayor09}), \cible\ should be considered  
for high-precision radial-velocity programs, despite its activity~level.

\begin{figure}[h] 
\begin{center}
\vspace{1cm}
\includegraphics[scale=0.53]{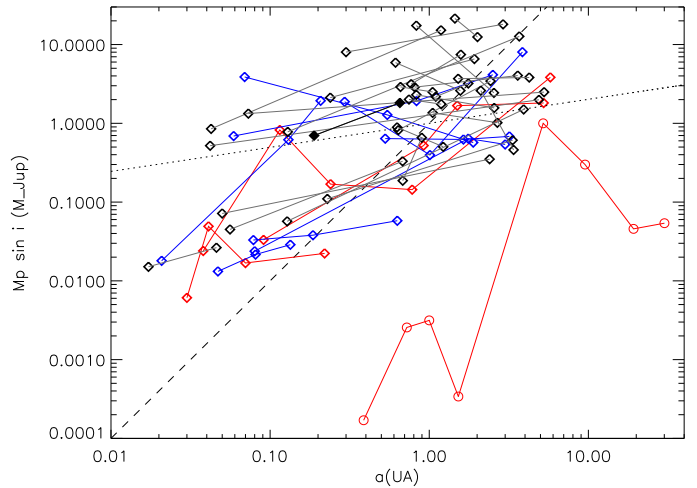}
\caption{Semi-major axes as a function of the projected masses for planets 
in multi-planet systems. 38 known extrasolar systems are plotted, with the planets of a given system 
that are linked by a solid line. The \cible-system is shown with filled diamonds. 
The fitted relation is plotted in dotted line (mass proportional 
to $a^{0.3}$); the dashed line show the $a^2$ relation. The 8 planets of the Solar System 
are also plotted for comparison (circles).
Systems with two, three and more planets are in black, blue and red, respectively.}
\label{fig_stat}
\end{center}
\end{figure}

The \cible\ system presents a hierarchical disposition, with the inner planet 
being the less massive one and the outer planet being the more massive. 
Fig.~\ref{fig_stat} displays the mass~--~semi-major axis relation for 
known multi-planetary systems. The data 
are taken from the compilation of the Extrasolar Planets 
Encyclopedia\footnote{http://exoplanet.eu}. Most of the known multi-planetary systems 
show such hierarchical disposition, as roughly the Solar System. The fitted relation 
between those two parameters provides a planet mass proportional to $a^{0.3}$ 
($\propto a^{1.5}$ for the Solar System). The plot suggests the slope could be deeper 
for systems including low-mass planets.  
A positive slope could be due to a 
higher migration efficiency for low-mass planets, and/or to the fact that giant 
planets are preferentially formed at 
larger distances of their host stars than low-mass planets. However, observational 
biases are important here, as low-mass planets are easier to detected at short 
orbital periods from radial velocity variations. The semi-amplitude of the reflex motion 
of a star due to a planetary companion 
is proportional to $\sqrt{a}\times M_\textrm{p} \sin i$, so one could expect a 
$a^{2}$-dependance in Fig.~\ref{fig_stat}. As the averaged slope is~lower, this could suggest 
there is actually no strong dependance on average between those two parameters for 
multi-planet~systems. 

Fig.~\ref{fig_stat} also shows that only a few multi-planetary systems include close-in giant 
planets. This agrees with Wright et al.~(\cite{wright09}), who reported that single-planet 
systems show a pileup at 3-day period and a jump at $a\simeq 1$~AU, while multi-planet 
systems show a more uniform distribution. Still, the close-in planets in known multi-planet 
system are mainly low-mass planets. Hot Jupiters appear to be sparse in multi-planet systems, 
showing here again a distribution which is different 
from single-planet systems. Only five hot-Jupiters are known to be in planetary systems, namely 
HIP\,14810b, ups~And\,b, HAT-P-13b,  HD\,187123b, and HD\,217107b. 
Single- and multi-planet systems thus appear to have significant differences 
in some of their properties. 
Such differences may provide clues allowing a better understanding of the formation 
and evolution of those systems.
Improving the statistic of extra-solar planets should be continued, 
in particular in multi-planet systems and with radial velocity~surveys.

\begin{acknowledgements}
We thank A.~C.~M.~Correia for helps and discussions, as well as 
all the staff of Haute-Provence Observatory for their 
support at the 1.93-m telescope and on \sophie.
We thank the ``Programme National de Plan\'etologie'' (PNP) of CNRS/INSU, 
the Swiss National Science Foundation, 
and the French National Research Agency (ANR-08-JCJC-0102-01 and ANR-NT05-4-44463) 
for their support to our planet-search programs.
NCS would like to thank the support by the European Research Council/European 
Community under the FP7 through a Starting Grant, as well from Funda\c{c}\~ao 
para a Ci\^encia e a Tecnologia (FCT), Portugal, through a Ci\^encia\,2007 contract 
funded by FCT/MCTES (Portugal) and POPH/FSE (EC), and in the form of grants 
reference PTDC/CTE-AST/098528/2008 and PTDC/CTE-AST/098604/2008 from~FCT/MCTES.
\end{acknowledgements}


\begin{thebibliography}{}

\bibitem[1996]{baranne96} 
Baranne, A., Queloz, D., Mayor, M., et al. 1994, \aaps, 119, 373

\bibitem[2009a]{boisse09a} 
Boisse, I., Moutou, C., Vidal-Madjar, A., et al. 2009a, \aap, 495, 959

\bibitem[2009b]{boisse09b} 
Boisse, I., Bouchy, F., Chazelas, B., Perruchot, S., Pepe, F., Lovis, C., H\'ebrard, G. 2009b, 
\textit{New technologies for probing the diversity of brown dwarfs and exoplanets}, 
EPJ Web of Conferences, in press

\bibitem[2005]{bouchy05} 
Bouchy, F., Pont, F., Melo, C., Santos,  N. C., 
Mayor, M., Queloz, D., Udry, S., 2005, \aap, 431, 1105

\bibitem[2009]{bouchy09} 
Bouchy, F., H\'ebrard, G., Udry, S., et al. 2009, \aap, 505, 853

\bibitem[1999]{chambers99} 
Chambers, J.E.  1999, \mnras, 304, 793

\bibitem[2007]{cochran07} 
Cochran, W. D., Endl, M., Wittenmyer, R. A., Bean, J. L. 2007, \apj, 665,1407

\bibitem[2005]{correia05}
Correia, A. C. M., Udry,  S., Mayor, M., Laskar,  J.,  Naef , D., Pepe,  F., Queloz, D., 
Santos,  N. C. 2005, \aap, 440, 751

\bibitem[2008]{dasilva08} 
Da Silva, R., Udry, S., Bouchy, F., et al. 2008, \aap, 473, 323

\bibitem[2004]{fernandes04}
Fernandes, J., Santos, N. C. 2004, \aap, 427, 607 

\bibitem[2008]{hebrard08} 
H\'ebrard, G., Bouchy, F., Pont, F., et al. 2008, \aap, 481, 52

\bibitem[2009]{hebrard09} 
H\'ebrard, G., Udry, S., Lo~Curto, G., et al. 2009, \aap, in press [arXiv:0912.3202]

\bibitem[2008]{loeillet08} 
Loeillet, B., Shporer, A., Bouchy, F., et al. 2008, \aap, 481, 529

\bibitem[2006]{lovis06} 
Lovis, C., Mayor, M., Pepe, F., et al. 2006, \nat, 441, 305

\bibitem[2004]{mcarthur04}
McArthur, B., Endl, M., Cochran W., et al. 2004, \apj, 614, L81

\bibitem[2008]{mamajek08} 
Mamajek, E. E., \& Hillenbrand, L. A. 2008, \apj, 687, 1264

\bibitem[2009]{mayor09} 
Mayor, M., Udry, S., Lovis, C., et al. 2009, \aap, 493, 639

\bibitem[2007]{melo07} 
Melo, C., Santos, N. C., Gieren, W., et al. 2007, \aap, 467, 721

\bibitem[2009]{moutou09}
Moutou, C., H\'ebrard, G., Bouchy, F., et al. 2009, \aap, 498, L5
 
\bibitem[1984]{noyes84}
Noyes, R. W., Hartmann, L. W., Baliunas, S. L., Duncan, D. K., Vaughan, A. H.	1984, \apj, 279, 763

\bibitem[2002]{pepe02} 
Pepe, F., Mayor, M., Galland, F., et al. 2002, \aap, 388, 632

\bibitem[2007]{pepe07} 
Pepe, F., Correia, A. C. M.., Mayor, M., et al. 2007, \aap, 462, 776

\bibitem[2008]{perruchot08}
Perruchot, S., Kohler, D., Bouchy, F., et al., 2008, in \textit{Ground-based and Airborn
Instrumentation for Astronomy II}, Edited by McLean, I.S., Casali, M.M., Proceedings of the 
SPIE, vol. 7014, 70140J

\bibitem[1997]{perryman97} 
Perryman, M. A. C., Lindegren, L., Kovalevsky, J., et al. 1997,  \aap, 323, L49

\bibitem[2009]{pont09}
Pont, F., H\'ebrard, G., Irwin, J. M., et al. 2009, \aap, 509, 695
 
\bibitem[2001]{queloz01}
Queloz, D., Henry, G. W., Sivan, J. P., et al. 2001, \aap, 379, 279

\bibitem[2000]{santos00} 
Santos, N. C., Mayor, M., Naef, D., et al. 2000,  \aap, 361, 265

\bibitem[2002]{santos02} 
Santos, N. C., Mayor, M., Naef, D., et al. 2002, \aap, 392, 215

\bibitem[2004]{santos04} 
Santos, N. C., Israelian, G., Mayor, M. 2004, \aap, 415, 1153

\bibitem[2005]{santos05} 
Santos, N. C., Israelian, G., Mayor, M., et al. 2005, \aap, 437, 1127 

\bibitem[2008]{santos08} 
Santos, N. C., Udry,  S., Bouchy, F., et al. 2008, \aap, 487, 369

\bibitem[2009]{wright09} 
Wright, J. T., Upadhyay, S., Marcy, G. W., Fischer, D. A., Ford, E. B., Johnson, J. A. 2009, 
\apj, 693, 1084

\end{thebibliography}
\end{document}